\documentclass[10pt]{emulateapj}

\usepackage{graphicx}

\newcommand{\is}{\ensuremath{\!=\!}}
\newcommand{\CaII}{Ca~II}
\newcommand{\CaIIH}{Ca~II~H}
\newcommand{\CaIIHK}{Ca~II~H\&K}
\newcommand{\CaIIIR}{Ca~II~854.2~nm}
\newcommand{\Halpha}{H$\alpha$}
\begin{document}

\title{Three-dimensional non-LTE radiative transfer computation of the
  Ca 8542 infrared line from a radiation-MHD simulation}

\author{J. Leenaarts\altaffilmark{1}, M. Carlsson\altaffilmark{1} V. Hansteen\altaffilmark{1}, L. Rouppe van der Voort\altaffilmark{1}}
\affil{Institute of Theoretical Astrophysics, University of Oslo,
  P.O. Box 1029 Blindern, N-0315 Oslo, Norway}
\email{jorritl@astro.uio.no}

\altaffiltext{1}{also at Center of Mathematics for Applications \\
 University of Oslo, P.O. Box 1053 Blindern, N-0316 Oslo, Norway}

\begin{abstract}
  Interpretation of imagery of the solar chromosphere in the widely
  used \CaIIIR\ infrared line is hampered by its complex,
  three-dimensional and non-LTE formation. Forward modelling is
  required to aid understanding.  We use a 3D non-LTE radiative
  transfer code to compute synthetic \CaIIIR\ images from a
  radiation-MHD simulation of the solar atmosphere spanning from the
  convection zone to the corona. We compare the simulation with
  observations obtained with the CRISP filter at the Swedish 1--m
  Solar Telescope. We find that the simulation reproduces dark patches
  in the blue line wing caused by Doppler shifts, brightenings in the
  line core caused by upward-propagating shocks and thin dark
  elongated structures in the line core that form the interface
  between upward and downward gas motion in the chromosphere. The
  synthetic line core is narrower than the observed one, indicating
  that the sun exhibits both more vigorous large-scale dynamics as
  well as small scale motions that are not resolved within the
  simulation, presumably owing to a lack of spatial resolution.

\end{abstract}


\section{Introduction} 
The solar chromosphere has traditionally been observed in the \CaIIHK\
and the \Halpha\ lines. The use of these lines as a diagnostic suffers
from significant drawbacks. The \CaIIHK\ lines have wavelengths in the
violet part of the spectrum. A trade-off has to be made between filter
width and exposure time due to the lack of photons and the lower
sensitivity of CCD chips at these wavelengths.  Either the filter has
to be wide, leading to significant low-chromospheric line wing
contributions, or the long exposure time leads to lower spatial
resolution due to difficulties with the image restoration
\citep[e.g.][]{1999ApJ...517.1013L,2006A&A...459L...9W}. 
In addition, the \CaIIHK\ lines are subject to partial redistribution
(PRD), which complicates proper modeling of these lines.

Observations in the \Halpha\ line suffer to a much lower extent from
these effects, but \Halpha\ line formation is very complex and proper
modeling requires non-equilibrium ionization of hydrogen to be taken
into account, physics that so far has only been treated in detail in
one-dimensional (1D) models
\citep{2002ApJ...572..626C}.

These problems are much less severe for the \CaIIIR\
line. Non-equilibrium and PRD effects are less important for its
formation
\citep{1989A&A...213..360U}.
Observations can be done with narrow filters and short exposure times,
yielding a clean and high resolution view of the chromosphere. This
makes it an excellent diagnostic, both from the observational as the
modeling point-of-use, albeit at a lower diffraction-limited
resolution than the \CaIIHK\ and the \Halpha\ lines.

Fabry-P\'erot instruments such as IBIS
\citep{2006SoPh..236..415C} 
and CRISP
\citep{2008ApJ...689L..69S} 
now routinely observe the \CaIIIR\ line with high spectral, spatial,
and temporal resolution. However, interpretation of data obtained with
these instruments is non-trivial; forward modeling by computation of
synthetic filtergrams from radiation-MHD simulation is needed to gain
insight in the formation of the line.

The formation of the \CaIIIR\ line has been studied by
\citet{2006ApJ...639..516U} 
in the 1D semi-empirical FALC model
\citep{1993ApJ...406..319F} 
and the 1D dynamic models of
\citet{1999AIPC..471...23C} 
and a 3D snapshot of convection by
\citet{2000A&A...359..743A}. 
The 3D model did not include magnetic fields and extended up to 1~Mm
above $\tau_{500}\is1$, lower than the formation height of the line
core in the FALC model 
\citep[see Fig.~5 of][]{2008A&A...480..515C}. 
In this paper we present synthetic images of the \CaIIIR\ line
computed with a 3D non-LTE radiative transfer code from a snapshot of
a radiation-MHD simulation that extends from the convection zone to
the corona, capturing the whole height range of formation of the line.

\section{Observations}

Figure~\ref{fig:images} shows observations in the \CaIIIR\ line
obtained with the CRISP imaging spectropolarimeter
\citep{2008ApJ...689L..69S}
at the Swedish 1--m Solar Telescope.  The data was obtained on June
13, 2008 and was taken at disk-center in a coronal hole.  A scan
through the line was made using 24 filter positions from $-0.92$ to
$+0.19$~\AA\ with a FWHM of the filter of 110~m\AA. The image quality
was improved by post-processing using the MOMFBD algorithm
\citep{2005SoPh..228..191V}. 

\section{Radiative transfer computation}

The radiative transfer computation was based on a snapshot computed
with the Oslo Stagger Code
\citep{2007ASPC..368..107H}. 
This code employs an LTE equation of state and includes non-LTE
radiative losses using a multi-group opacity method and thin radiative
cooling in the corona and upper chromosphere. It includes thermal
conduction along magnetic field lines. The simulation had $256 \times
128 \times 160$ grid points, corresponding to a physical size of $16.6
\times 8.3 \times 15.5$~Mm. The snapshot has an averaged unsigned
magnetic field strength of 150~G and covers a height range from 1.5~Mm
below the photosphere up to 14~Mm above it, spanning from the upper
convection zone up into the corona. For the radiative transfer
computation the grid was interpolated onto a grid with 213~ points in
the $z$-direction covering a height range between -0.5 and 5.3~Mm
relative to $\tau_{500} \is 0$. The \CaII\ model atoms consists of 5
bound levels plus a continuum with 5 lines (\CaIIHK\ and the infrared
triplet) and 5 bound-free transitions. The \CaIIIR\ line had 200
frequency points. All lines were computed in complete
redistribution. Velocity fields were taken into account and no
additional microturbulence was added. The electron density was
computed assuming LTE ionization for all relevant
species. Photoionization by hydrogen Lyman lines was not taken into
account. The ray quadrature was computed using the A2 set of
\citet{carlson1963}.
The radiative transfer computation was performed using an
MPI-parallelized, domain-decomposed code named MULTI3D. This code is
based on the 1D code MULTI
\citep{1985JCoPh..59...56S}
and includes the same physics, but has a 3D short-characteristics
formal solver, allowing evaluation of the 3D radiation field.

\section{Results}

\begin{figure}
 \includegraphics[width=\columnwidth]{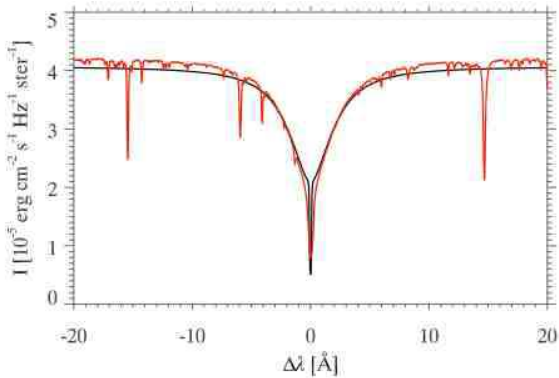}
 \caption{Average profile of the Ca 8542 IR line.  Solid red: FTS
   atlas; solid black: simulation. The simulated continuum has a lower
   intensity and the line core is narrower than in the
   observed profile. \label{fig:ftscomp}}
\end{figure}

Figure~\ref{fig:ftscomp} compares the average quiet sun profile of
the \CaIIIR\ line from the FTS atlas 
\citep{1984SoPh...90..205N} 
to the average profile computed from the simulation. The simulated
continuum is 3.5\% less bright than the observations, the line wings
fit rather well, while the line core is narrower and the bisector does
not show the inverse-C shape of the observations.

\begin{figure}
 \includegraphics[width=\columnwidth]{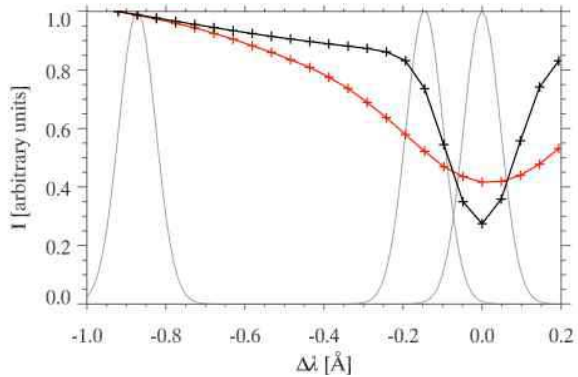}
 \caption{Line core of the Ca 8542 IR line. The profiles have been
   normalized on their values at $\Delta \lambda \is -0.92$~\AA. Solid
   red: CRISP; solid black: simulation; plus signs indicate the
   different filter positions. The gray curves show the shape and
   positions of the filter used to produce the synthetic images in
   Fig.~\ref{fig:images}. The synthetic line core is narrower than the
   observed one. \label{fig:profiles}}
\end{figure}

\begin{figure*}
  \includegraphics[width=\textwidth]{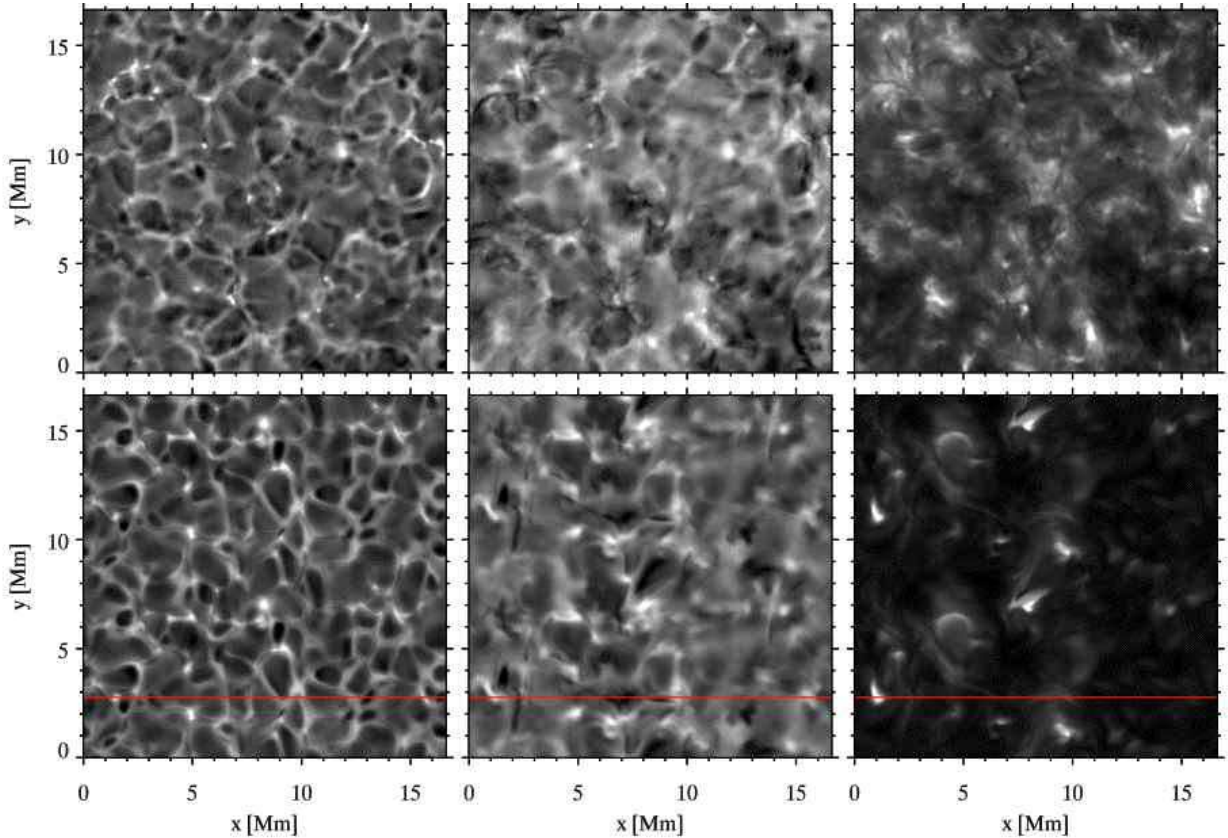}
  \caption{Observed (top row) and synthetic (bottom row) \CaIIIR\
    images at different positions in the line. Left: at $\Delta
    \lambda \is -0.87$\,\AA; middle: in the ''knee'' of the line, for
    the observations at $\Delta \lambda \is -0.39$\,\AA\ and for the
    synthetic image at $\Delta \lambda \is -0.14$\,\AA; right: in the
    line core at $\Delta \lambda \is 0$\,\AA. The brightness scales
    are clipped at the brightest and darkest 0.1\% of the pixels. The
    observed and synthetic images show similar structure in all panel
    pairs. The wavelengths of the middle panel pair are in the ''knee'' of
    the line profile, and because of the difference in the line core width
    this is not at the same wavelength. The red line indicates the cut
    displayed in Figs.~\ref{fig:contfunc} and~\ref{fig:spectrum}.
    \label{fig:images}}
\end{figure*}

The vertically emergent intensity in the \CaIIIR\ line was convolved
with the CRISP filter transmission function. Figure~\ref{fig:profiles}
compares the observed and synthetic profile averaged over the
simulation box and a quiet region of the same size in the observations
(shown in Fig.~\ref{fig:images}). The synthetic profile has a less
steep inner wing and a narrower line core.

\begin{figure*}
  \includegraphics[width=\textwidth]{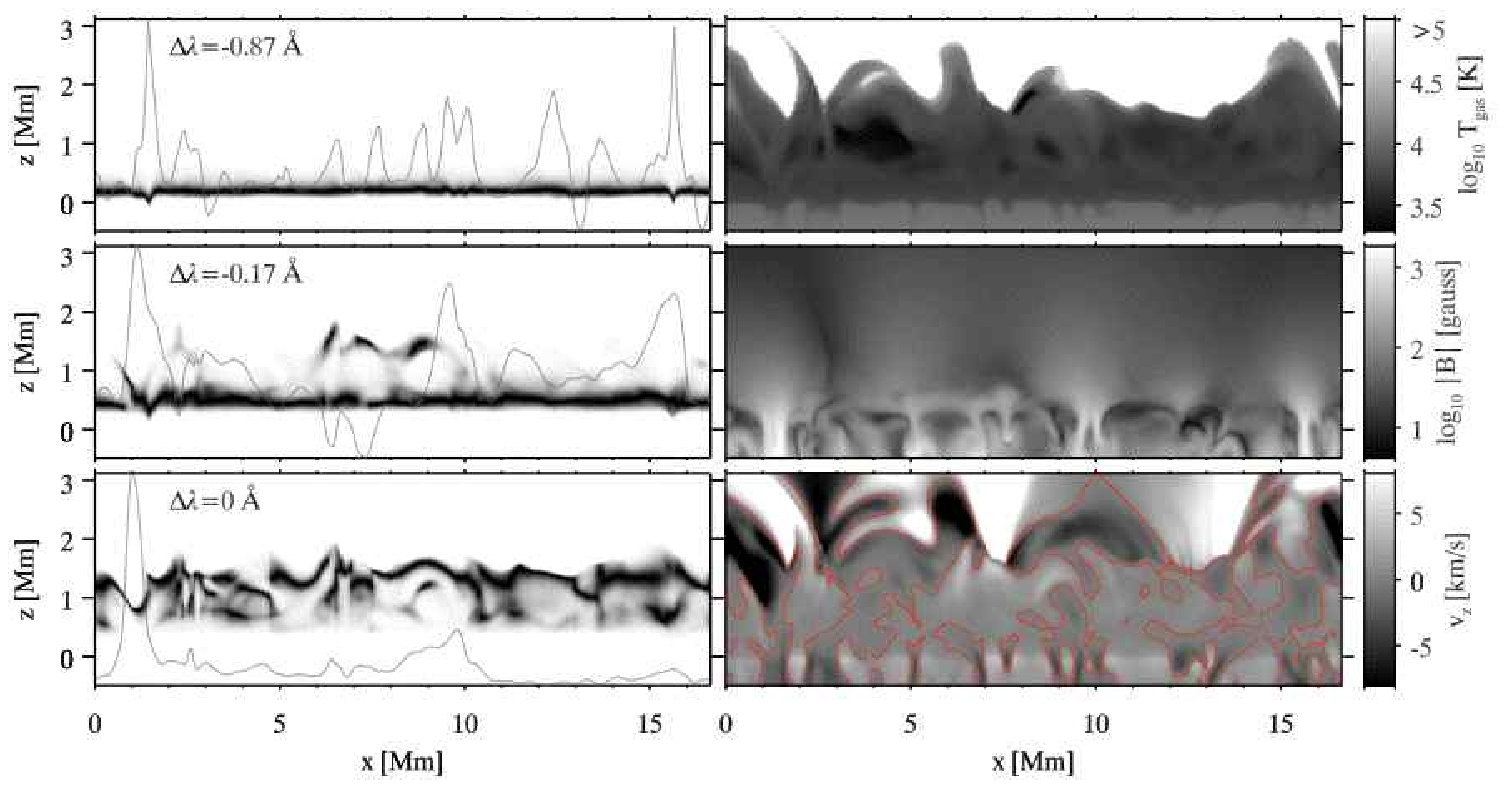}
  \caption{Vertical cut through the atmosphere along the red line in
Fig.~\ref{fig:images}. The left-hand panels show the contribution
function to the intensity within the CRISP filter, on an inverted
brightness scale and with each column scaled to maximum contrast for
improved visibility. The overplotted grey curves indicate the emergent
intensity in arbitrary units. The right-hand panels show, from top to
bottom, the gas temperature, the absolute magnetic field strength and 
the vertical velocity (positive means upflow). The red curve in the 
velocity panel indicates the zero vertical velocity contour.
\label{fig:contfunc}}
\end{figure*}

Figure~\ref{fig:images} compares a quiet sun region from the
observations to the synthetic filtergrams. The left panels show
reversed granulation with several magnetic bright points. The
contribution function in the upper-left panel of
Fig.~\ref{fig:contfunc} shows that the intensity in the magnetic
bright points are formed deeper than in the surrounding
low-field-strength atmosphere, similar to G~band bright points
\citep{2004ApJ...610L.137C}. 

The middle panels of Fig.~\ref{fig:images} show the atmosphere as seen
in the ''knee'' of the line, where the profile changes from the LTE
wings to the non-LTE formed line core
\citep{2008A&A...480..515C}.
Because of the difference in the line widths in the simulations and
the observations this is at different offsets relative to line center,
at $-0.39$\,\AA\ for the observations and $-0.14$\,\AA\ for the
simulations. Nevertheless, they show a remarkably similar scene. The
bright background shows the upper photosphere and lower chromosphere
at around $z \is 0.5$\,Mm where the reversed granulation pattern
changes to one dominated by acoustic shock interference
\citep{2004A&A...414.1121W}.

\begin{figure}
  \includegraphics[width=\columnwidth]{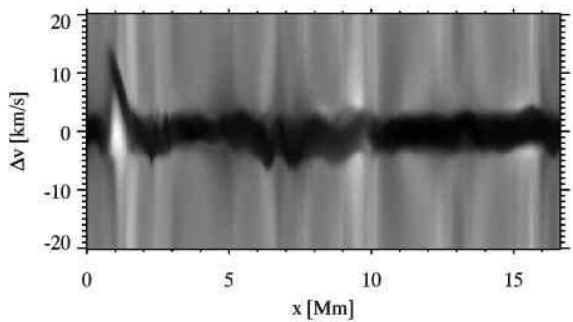}
  \caption{Emergent spectrum of the \CaIIIR\ line along the cut indicated in
    Fig.~\ref{fig:images}, with positive velocity corresponding to a
    redshift. The brightening at $(x,y) \is (1,3)$\,Mm in the
    lower right panel of Fig.~\ref{fig:images} is associated with a 12\,km/s
    redshift of the line core. The dark patches between $y \is 5$
    and $y \is 10$~Mm along the cut in the lower middle panel of
    Fig.~\ref{fig:images} are associated with a blueshift of 5\,km/s
    \label{fig:spectrum}}
\end{figure}

The upper-middle panel of Fig.~\ref{fig:images} shows black
structures, for example at $(15,1)$ and at $(15.5,13)$\,Mm. Similar
roundish or thin elongated structures appear in the synthetic image.
The corresponding contribution function panel in
Fig.~\ref{fig:contfunc} shows that they are formed at around $z \is
1.5$~Mm, 1\,Mm higher than the bright background. Comparison with the
lower right panel shows that the dark patches are associated with
upflows and the line profile in Fig.~\ref{fig:spectrum} shows a
blueshifted line at the positions of the dark patches.

There are three brightenings along the cut in the lower middle panel
of Fig.~\ref{fig:images}, at $x \is 1$, $x \is 9.5$ and $x \is
15.5$\,Mm, all located above a photospheric magnetic field
concentration (middle right panel of Fig.~\ref{fig:contfunc}).

The right panels show the line core. The simulated image has a higher
contrast, which can at least partly be explained by the fact that the
observations are not corrected for scattered light. The observations
show an amorphous background with a number of brightenings, such as at
$(x,y) \is (3.5,3.5)$ and $(16,8)$\,Mm. They are often, but not
always, located on top of or near bright points in the
photosphere. The simulated line core image (lower right) shows similar
brightenings, for example at $(x,y) \is
(1,3)$\,Mm. Fig.~\ref{fig:contfunc} shows that this brightening is
associated with a strong downflow above a magnetic element.  This
brightening is formed much deeper than is the intensity along the rest
of the cut. A similar brightening at $(x,y) \is (9.5,1.5)$\,Mm is not
associated with a photospheric magnetic field concentration, however.
The corresponding line profile in Fig.~\ref{fig:spectrum} shows an
emission peak slightly bluewards of line-center and an intensity dip
on the red side of the line typical of an upward propagating wave with
downflowing material above it
\citep[see][for a detailed description in the case of the \CaIIH 2V
grains]{1997ApJ...481..500C}.
These shocks can be either acoustic or magneto-acoustic, the radiative
transfer mechanism causing them to be bright in the line core remains
the same, and a single snapshot alone is not sufficient to distinguish
between them. Analysis of a time-series is needed to determine the
wave type.

On top of the background in the upper right panel of
Fig.~\ref{fig:images} one can see filamentary dark structures that can
be both straight and curved. They are mainly concentrated in the top
left part of the image, but can be found throughout the field of
view. The simulated line-core image shows similar dark filamentary
structures (for example at $(x,y) \is (2.5,3)$ and $(10,6)$\,Mm). They
appear similar to the observed ones, but tend to be straighter, though
a slightly curved example can be found at $(x,y) \is (4,5)$\,Mm. They
are the interface between a patch of upflowing and downflowing
material. They form where the vertical velocity at their formation
height is zero and the line core is located at the rest frequency of
the line. On either side of the dark filament the gas is moving up or
down, shifting the line core to the blue or red and thus showing the
brighter line wings at the rest frequency.

\section{Discussion \& Conclusions}

The simulation reproduces a number of observed features rather well,
but there are a number important differences that carry large
diagostic potential.

The difference in the continuum intensities in Fig.~\ref{fig:ftscomp}
can be explained by a small difference in the effective temperature of
the simulation and the sun combined with the effect of box
oscillations in the simulation, which causes fluctuations in the
continuum intensity with an amplitude of about 1\%.

The simulated spatially-averaged line-core profile is narrower than
the observed one (Fig.~\ref{fig:profiles}). We computed the deviations
of the position of the line core from the rest-wavelength over the
field-of-view for both the simulation and the observation. The
standard deviation for the simulations is 1.1\,km/s, which is 50\%
smaller than for the observations. This indicates that the solar
chromosphere has more vigorous large-scale dynamics than the
simulation. However, this effect is too small to explain the
difference in average line-core width.

We also inspected the line-core widths for individual resolution
elements. The simulated line cores are on average narrower than the
observed ones. This is the dominant reason for the difference in the
average core width, indicating that the sun is dynamic on scales not
resolved by the simulation.

Close inspection of the line core images of Fig.~\ref{fig:images}
shows that the observations have more structure on scales of about
0.1~Mm than the simulation, also hinting that the simulation does not
have sufficient resolution to generate enough small-scale motion to
broaden the line core.


In summary, we have computed the 3D, non-LTE, emergent intensity in
the \CaIIIR\ line from a snapshot of a radiation-MHD simulation. It
reproduces a number of observed features rather well, i.e., black
patches in the blue 'knee' of the line, and thin dark lines and
brightness enhancements in the line core. However, the simulated line
core is narrower than the observed one, indicating that the sun shows
more vigorous large scale dynamics and motions at smaller scales than
can be resolved in the simulation. We intend to repeat the same
analysis with a radiation-MHD simulation with higher spatial
resolution.

\acknowledgments J. Leenaarts is supported by the European Commission
funded Research Training Network SOLAIRE. This research was supported
by the Research Council of Norway through grant 170935/V30 and a grant
of computing time from the Program for Supercomputing. The Swedish 1-m
Solar Telescope is operated on the island of La Palma by the Institute
for Solar Physics of the Royal Swedish Academy of Sciences in the
Spanish Observatorio del Roque de los Muchachos of the Instituto de
Astrof{\'\i}sica de Canarias.





\end{document}